\documentstyle[aps,prb,epsf]{revtex}
\begin{document}
\input psfig.sty
\draft

\title{On the value of the Curie temperature in doped manganites}
\author{M. O. Dzero}
\address{National High Magnetic Field Laboratory, Florida State University,
Tallahassee, FL, 32304, USA}
\date{\today}
\maketitle

\begin{abstract}

We have verified that the variational mean field theory approach 
suggested by Narimanov and Varma (preprint cond-mat/0002191) 
being applied to the realistic two-band model provides a
good agreement with experimental data for the Curie temperature in
doped manganites A$_{1-x}$B$_x$MnO$_3$ ($x\simeq{0.3}$).
We have also considered the problem of an interplay between 
the ferromagnetic and antiferromagnetic interactions by using
the same approach.

\end{abstract}
The problem of finding the value of the critical temperature
for the paramagnetic-to-ferromagnetic transition
in doped manganites has presented a lot of interest recently. 
An agreement of theoretical predictions with experimental 
results for the Curie temperature may shed some light on 
physical mechanisms driving the transition near the 
concentration region, where the colossal magneto-resistance (CMR) 
effect takes place. 

An interesting approach for obtaining the value 
for the transition temperature has been recently proposed 
by Narimanov and Varma \cite{Narimanov} 
(see also \cite{Varma}) and is based on the variational mean field 
theory within the formalism of the double exchange (DE) model 
\cite{Zener,Anderson,deGennes}. 
The final result for the critical temperature dependence on
concentration \cite{Narimanov,Varma} can be approximately written 
in the following form:
\begin{eqnarray}
T_c{~\simeq}~x~(1-x)~W,
\label{curie1}
\end{eqnarray}
where $W$ is a bandwidth. For $W=1.8$ eV and $0.15\leq{x}\leq{0.3}$ 
Eq. (\ref{curie1}) recovers the experimentally observed values for the 
Curie temperature. This result can be generalized in such a way,
that the DE model works well in order to describe the physics
near the CMR region \cite{Furukawa,Nagaev} and in principle
one does not need to take into account 
other mechanisms \cite{Millis1,Millis2} 
which might effect the transition. 

In what follows we adopt the variational mean field theory scheme 
for the DE model via degenerate orbitals \cite{Khomskii} in order 
to verify the validity of the two-band approach 
\cite{Gor'kov,Dzero} near the CMR region.
In the DE model within the two-band scheme, 
there is no electron-hole symmetry, which means that
the Curie temperature is non-symmetric with respect to
substitution $x\rightarrow{1-x}$. Since for the small concentrations 
percolation effects play an essential role \cite{Gor'kov,Gorkovrev},
we will be primarily concerned with obtaining the value for 
the Curie temperature near the CMR region ($x\sim{0.3}$).
As we will see, the approach we are using works well and recovers
the right numerical value for the Curie temperature in units
of hopping amplitude the value for which in the two-band
model for manganites has been estimated in \cite{Dzero}.
The relative simplicity of present approach also allows us
to consider a problem of suppression of the ferromagnetic
transition by the superexchange interaction \cite{Yi,Okamoto,Alonso}. 

We consider three $t_{2g}$ electrons on a Mn ion as the 
localized classical spins. The Hamiltonian for the DE model on
the degenerate orbitals can be defined as:
\begin{eqnarray}
H_{DE}&=&-\sum\limits^{}_{\langle{ij}\rangle}t^{ab}_{ij}
{\cdot}c^{\dagger}_{ia\alpha}c_{jb\alpha}~-~J_H\sum\limits^{}_{i}
\vec{S_i}{\cdot}
c^{\dagger}_{ia\alpha}\vec{\sigma}_{\alpha\beta}c_{ia\beta}.
\end{eqnarray}
The matrix elements $t^{ab}_{ij}$ describe the electrons, hopping 
from one site to another, on two-fold degenerate $e_g$ orbitals.
As it was shown by Anderson and Hasegawa ~\cite{Anderson}, in the
limit of strong Hund's coupling, the hopping amplitude of the electrons
between the sites $i$ and $j$ acquires an angular dependence 
$t_{ij}=t(\theta_{ij})$, 
where $\theta_{ij}$ is the angle between the core
spins $\vec{S}_i$ and $\vec{S}_j$. In the DE model the hopping amplitude
is the largest, when the spins are aligned, 
which means that a ferromagnetic
state minimizes the kinetic energy. Thus, the effective Hamiltonian for
our problem can be written as:
\begin{eqnarray}
H&=&H_{DE} + 
J\sum\limits^{}_{\langle{ij}\rangle}{\vec{S}_i{\cdot}\vec{S}_j}.
\end{eqnarray}

We employ the variational mean field theory by introducing the 
spin orientation distribution function 
$P(\Omega)=\frac{1}{2\pi}P(\vartheta)$.
In the mean field approximation, the distribution function depends 
only on single spin orientation, which is described by an angle $\vartheta$.
The idea is to represent the free energy of our system as a functional of 
$P(\vartheta)$ and, following the standard mean field treatment, to find
the self-consistent equation for the distribution function.

The free energy is defined as:
\begin{equation}
F = E_{el} + E_{spin} - T\cdot{S_{spin}},
\end{equation}
where $S_{spin}$ is the entropy of our system, $E_{el}$ is electron energy,
$E_{spin}$ is the spin energy. 

As mentioned above, 
the distribution function for the system of spins is just the product 
of the distribution functions of the individual spins. 
Thus, in the semiclassical
limit,  the entropy can be written as:
\begin{eqnarray}
S_{spin}&=&-\int\limits^{\pi}_{0}d{\vartheta}\sin\vartheta P(\vartheta)\log
[P(\vartheta)].
\label{entropy}
\end{eqnarray} 

In the absence of a magnetic field, the spin energy contribution, 
$E_{spin}$, in the classical limit, is just
the average of the superexchange Hamiltonian:
\begin{eqnarray}
E_{spin}&=&\frac{J}{2}
\int\limits^{2\pi}_{0}\frac{d\phi_1}{2\pi}
\int\limits^{2\pi}_{0}\frac{d\phi_2}{2\pi}
\int\limits^{\pi}_{0}{d{\vartheta}_1}{\sin\vartheta_1}P(\vartheta_1)
\int\limits^{\pi}_{0}{d{\vartheta}_2}{\sin\vartheta_2}P(\vartheta_2)
\cos(\theta),
\label{spins}
\end{eqnarray}
where $\cos(\theta)=\cos\vartheta_1\cos\vartheta_2 + 
\sin\vartheta_1\sin\vartheta_2\cos(\phi_1-\phi_2)$.

The electron energy is calculated as follows. First, we calculate the energy
for a given spin distribution, $E[t]$ and then we just take the average
of $E[t]$ over all possible spin configurations:
\begin{eqnarray}
E_{el}&=&\int{dt}P(t)E[t].
\end{eqnarray}

To calculate $E[t]$ we adopt the tight binding approximation. 
For the two-dimensional representation $e_g$ we choose the normalized
complex functions $\psi_1$ and $\psi_2$, given by \cite{Gor'kov}:
\begin{eqnarray}
\psi_1\propto z^2+\epsilon x^2+\epsilon^2y^2, ~\psi_2\equiv\psi_1^{\ast},
\end{eqnarray}
where $\epsilon =\exp (2\pi i/3)$. 

The cubic spectrum consists of two branches:
\begin{eqnarray}
{\varepsilon}_{1,2}({\bf p})&=&-t(\theta)\cdot\left \{(c_x+c_y+c_z) \pm
\sqrt{c_x^2 + c_y^2 + c_z^2 - c_x c_y - c_y c_z - c_z c_x}~\right \}
\label{bands}
\end{eqnarray}
(we introduced the notations $c_i = \cos(k_i a), i=x,y,z$). 
Since there is an interaction between
the $e_g$ electrons and the magnetic background,following the
discussion by Anderson and Hasegawa, we have introduced the effective
hopping amplitude $t(\theta)=\mid{A}\mid\cos(\theta/2)$ (${\theta}$
is the angle between the two neighboring spins and $\mid{A}\mid$ is
a hopping amplitude, whose numerical value will be 
defined later).

Thus, the electron energy for a given spin configuration is:
\begin{eqnarray}
E[t]&=&\sum\limits^{2}_{i=1}\int\frac{d{\bf k}}{(2\pi)^3}
\varepsilon_i({\bf k},\theta){\cdot}n_F[\varepsilon_i({\bf k},\theta)],
\label{kinetic}
\end{eqnarray}
where $n_F$ is Fermi distribution function.

Using the expressions (\ref{entropy},\ref{spins},\ref{kinetic}) and
the Lagrange variational method, the
free energy functional acquires the following form:
\begin{eqnarray}
&&F[P_{\vartheta};\mu, \xi]=
\int\limits^{2\pi}_{0}\frac{d\phi_1}{2\pi}
\int\limits^{2\pi}_{0}\frac{d\phi_2}{2\pi}
\int\limits^{\pi}_{0}{d{\vartheta}_1}{\sin\vartheta_1}P(\vartheta_1)
\int\limits^{\pi}_{0}{d{\vartheta}_2}{\sin\vartheta_2}P(\vartheta_2)
\times\nonumber\\
&&\left\{ \sum\limits^{2}_{i=1}\int\frac{d{\bf k}}{(2\pi)^3}
(\varepsilon_i({\bf k},\theta)-\lambda)
{\times}n_F[\varepsilon_i({\bf k},\theta)] +
\frac{J}{2}\cos(\theta)\right \} + 
T\int\limits^{\pi}_{0}d{\vartheta}\sin\vartheta\times\nonumber\\
&&P(\vartheta)\log[P(\vartheta)] + 
\xi\int\limits^{\pi}_{0}d{\vartheta}\sin\vartheta 
P(\vartheta) - \mu_BSB\int\limits^{2\pi}_{0}\frac{d\phi_1}{2\pi}
\int\limits^{\pi}_{0} d\vartheta_1\times\nonumber\\
&&\sin\vartheta_1\cos\vartheta_1
P(\vartheta_1) + \widetilde{F}[\lambda,\xi]
\label{freeenergy}
\end{eqnarray}
($\lambda$ is a Lagrange multiplier, being a constraint to have a
constant number of conduction electrons in our system, 
$\xi$ is a Lagrange multiplier, which provides the 
normalization for $P(\vartheta)$ and $\widetilde{F}[\lambda,\xi]$ is
a $\vartheta$ independent part of the free energy, $B$ is a external 
magnetic field). 
Since the number of conduction electrons is 
conserved in our system, the free energy functional 
acquires its minimal value
for $\lambda$ being equal to the chemical potential $\mu$, which
is defined, in our case, by the following expression:
\begin{eqnarray} 
1-x&=&
\sum\limits^{2}_{i=1}\int\frac{d{\bf k}}{(2\pi)^3}
n_F[(\varepsilon_i({\bf k},\theta=0)-\mu)/T],
\nonumber
\end{eqnarray}
where $x$ is the hole concentration.

Now, taking the functional derivative of (\ref{freeenergy}) with
respect to $P(\vartheta)$, we obtain the following integral equation:
\begin{eqnarray}
&&P(\vartheta)=\exp\left[-2 
\int\limits^{2\pi}_{0}\frac{d\phi_1}{2\pi}
\int\limits^{2\pi}_{0}\frac{d\phi_2}{2\pi}
\int\limits^{\pi}_{0}{d{\vartheta}_1}{\sin\vartheta_1}P(\vartheta_1)
\right.\times\nonumber\\
&&\left.\left\{ \sum\limits^{2}_{i=1}\int\frac{d{\bf k}}{(2\pi)^3}
\frac{(\varepsilon_i({\bf k},\theta)-\mu)}{T}
n_F[\varepsilon_i({\bf k},\theta)] +
\frac{J}{2T}\cos(\theta)\right \} - \nonumber\right.\\
&&\left.\xi/T + \frac{\mu_BSB}{T}\cos\vartheta \right ].
\label{integral}
\end{eqnarray} 

The integral equation (\ref{integral}) is nothing but {\it variational} 
mean-field theory equation for the distribution function $P(\vartheta)$, 
where $P(\vartheta)$ plays the role of the order parameter.
If the system is in the paramagnetic phase, 
$P(\vartheta)$ is uniform. Equation (\ref{integral}) allows a
numerical solution. Its results are shown on Fig. 1.
\begin{figure}[h]
\hspace{-1cm}
\centerline{\psfig{file=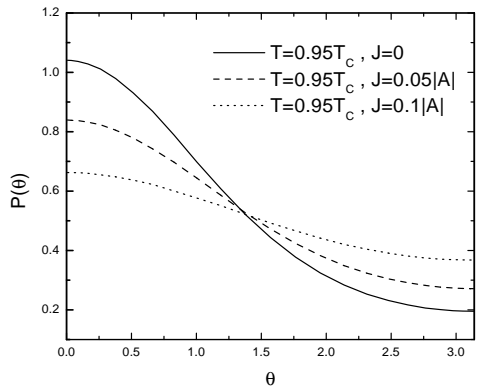,height=6cm,width=8cm}}
\caption{The single spin orientation distribution function $P(\vartheta)$ 
is shown as a function of $\vartheta$ for different values of the 
antiferromagnetic exchange constant $J$. The results are obtained
for hole concentration $x\simeq{0.3}$.} 
\end{figure}
As soon as the temperature decreases 
(in the presence of a small magnetic field) and approaches
the Curie temperature $T_c$, small deviations
from the uniform distribution will appear in $P(\vartheta)$:
\begin{eqnarray}
P(\vartheta){\mid}_{T\to{T_c}}&=&\frac{1}{2} + \Delta(\vartheta), 
~~\Delta(\vartheta)\ll{1}.
\label{delta}
\end{eqnarray}  
Since $\Delta(\vartheta)$ is small, we can expand both sides of Eq.
(\ref{integral}) in powers of $\Delta(\vartheta)$, keeping only 
the linear terms and solving the final expression for $\Delta(\vartheta)$.
The results can be written in the following form:
\begin{eqnarray}
\Delta(\vartheta)&{\propto}&\frac{B\cos\vartheta}{T-T_c},\nonumber\\ 
T_c&=&\int\limits^{\pi}_{0}d{\vartheta}{~\sin\vartheta}{~\cos\vartheta}
\left\{ \sum\limits^{2}_{i=1}\int\frac{d{\bf k}}{(2\pi)^3}
(\mu - \varepsilon_i({\bf k}))
{\times}n_F[\varepsilon_i({\bf k})] -
\frac{J}{2}\cos(\vartheta)\right \}.
\label{curie}
\end{eqnarray}
Thus, as it is seen from (\ref{curie}), the ferromagnetic transition is
linearly suppressed by the superexchange interaction, in agreement
with previous results obtained in \cite{Yi}. 
The energy scale is defined in the
units of the hopping amplitude, $\mid{A}\mid$. The experimental results
for the spin-wave stiffness coefficient provide $\mid{A}\mid$=0.16 eV
(see e.g. \cite{Dzero} and references therein). 

As we see in Fig. 2, $T_c$ decreases as system is doped with holes.
For concentrations, where the CMR effect is very pronounced 
($x\simeq 0.3$), the value we found for the critical temperature
is equal to $390$ K, which is in good agreement with the experimental 
data. Hence, the DE model via degenerate orbitals 
alone gives a good value for the 
critical temperature. For small concentrations our result does not 
agree with experimental observations, in which $T_c$ is increasing 
with doping.
Such a behavior for small concentrations can be explained
in terms of percolation theory (see \cite{Gor'kov,Dzero,Gorkovrev} 
and references therein) and can not be captured by using the 
present approach. 
\begin{figure}[h]
\hspace{-1cm}
\centerline{\psfig{file=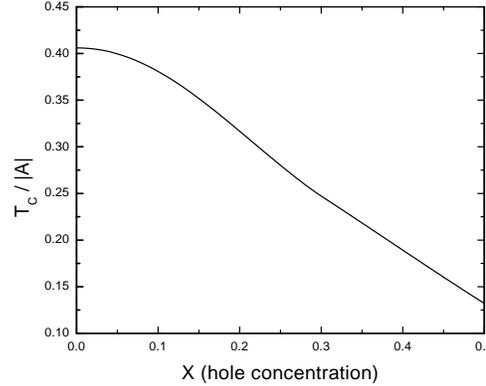,height=6cm,width=8cm}}
\caption{Curie temperature  
is plotted as a function of hole concentration $x$.}
\end{figure}
In conclusion, we have shown, that the variational mean field theory 
can successfully describe the paramagnetic-to-ferromagnetic 
transition for $x\sim{0.3}$ in the frame of the DE model 
when the two-fold degeneracy of the $e_g$ levels is taken 
into account. We have also discussed the competition 
between the ferromagnetic and antiferromagnetic exchange interactions.
Although the phonons might play an essential role in the high-temperature
properties of the manganites near the metallic region (isotope effect), 
the other mechanism for suppressing the ferromagnetic phase 
is the superexchange interaction between the $t_{2g}$ core spins.

We thank L. P. Gor'kov for very useful discussions and 
F. Drymiotis for help with the manuscript.
This work was supported by the National High Magnetic
Field Laboratory through NSF Cooperative Agreement \# DMR-9527035 and the
State of Florida.

\end{document}